\documentclass[a4paper,12pt]{article}

\usepackage{amsmath}
\usepackage{latexsym}
\usepackage{amsfonts}
\usepackage{amsthm}
\usepackage{bm}
\newtheorem{thm}{Theorem}[section]

\newtheorem{rem}[thm]{Remark}

\newtheorem{ex}{Example}
\title{Nonlocal Hydrodynamic Type of  Equations}

\author{Metin G{\" u}rses \thanks{
email: gurses@fen.bilkent.edu.tr} \\
{\small Department of Mathematics, Faculty of Sciences}\\
 {\small Bilkent University, 06800 Ankara, Turkey}~~~~\\
 Asl{\i} Pekcan \thanks{
email: aslipekcan@hacettepe.edu.tr}\\
{\small Department of Mathematics},\\
{\small Hacettepe University, 06800 Ankara, Turkey}~~~~\\
 Konstyantyn Zheltukhin \thanks{
email: zheltukh@metu.edu.tr}\\
{\small Department of Mathematics},\\
{\small Middle East Technical University, 06800 Ankara, Turkey}}

\date{\nonumber}
\begin{document}

\maketitle

\date{\nonumber}

\begin{abstract}
We show that the integrable equations of hydrodynamic type  admit nonlocal reductions. We first construct such reductions for a general Lax equation and then give several examples.
The reduced nonlocal equations are of hydrodynamic type and integrable. They admit Lax representations and hence possess infinitely many conserved quantities.

\vspace{0.3cm}
\end{abstract}

%\pacs{}
% insert suggested keywords - APS authors don't need to do this

%\maketitle must follow title, authors, abstract, \pacs, and \keywords
\maketitle

\section{Introduction}

One way of obtaining a system of integrable equations, in Gelfand-Dikii formalism \cite{GD}, is to construct a Lax operator on some Lie algebra.
The examples of such algebras are the matrix algebra,  algebra of differential operators, and algebra of Laurent series (see \cite{Li}-\cite{Bl} and references there in). In most cases the Lax operator is polynomial. It is a polynomial of the spectral parameter in the matrix algebra, the operator $D_{x}$ in the  algebra of differential operators, and an auxiliary variable $p$ (momentum) in the case of the algebra of Laurent series.

Writing the Lax equation on the algebra of Laurent series we obtain  equations of the hydrodynamic type. These equations are integrable. This means that they admit  recursion operators, multi-Hamiltonian representation (see \cite{Li}, \cite{GurZhel}-\cite{Zhel2}). They also have many types of reductions.

For a hydrodynamic type system let us give an example
\begin{equation}\label{system}
\begin{array}{l}
  q^i_t=F(q^j,q^j_x,r^j,r^j_x), \qquad i,j=1,\cdots, L,  \\
  r^i_t=G(q^j,q^j_x,r^j,r^j_x), \qquad i,j=1,\cdots, L,  \\
 \end{array}
\end{equation}
where $F,G$ are functions of the dynamical variables $q^{i}(t,x)$, $r^{i}(t,x)$,  and their first partial derivatives with respect to $x$.
We define  a reduction as the following relations
\begin{equation} \label{red1}
 r^i(x,t)=\rho q^i(\varepsilon_1 x,\varepsilon_2 t), \quad i=1,\cdots, L,
\end{equation}
where $\varepsilon_1, \,\varepsilon_2,\, \rho$ are  real constants such that  $\rho^2=\varepsilon_1^2=\varepsilon_2^2=1$.
In the case of complex valued dependent variables we also consider the following relations
\begin{equation}\label{red2}
 r^i(x,t)=\rho \bar q^i(\varepsilon_1 x,\varepsilon_2 t), \quad i=1,\cdots, L.
\end{equation}
 The system \eqref{system} must remain consistent with respect to the reductions (\ref{red1}) and (\ref{red2}).

If $\varepsilon_1=\varepsilon_2=1$ we have a local reduction. Such reductions were considered in \cite{GurZhel}. If $\varepsilon_1=-1$ or $\varepsilon_2=-1$ we have nonlocal reductions. In this work we address to the problem of nonlocal reductions for the Lax equations on the algebra of Laurent series.

The nonlocal reductions of systems of integrable equations were first consistently applied to the nonlinear Schr\"odinger system, which led to the nonlocal nonlinear Schr\"odinger equation (nNLS) \cite{AbMu1}-\cite{AbMu3}. It has been later shown that there are many other systems where the consistent nonlocal reductions are possible \cite{fok}-\cite{Pek}. For a recent review of this subject see \cite{GurPek3}. See also \cite{super} for the discussion of superposition of nonlocal integrable equations and \cite{origin} for the origin of nonlocal reductions.

The layout of the paper is as follows. In Section 2 we give a short review on the hydrodynamic type of (dispersionless) equations and their Lax representations. In Section 3 we give nonlocal reductions for real and complex valued fields for $N=2$. We obtain some explicit examples of nonlocal hydrodynamic type of equations with some conserved quantities. In Section 4 we discuss the nonlocal reductions in general and give examples for higher values of $N$.

\section{Lax equations}

In this section we describe the algebra of Laurent series, introduce necessary definitions to write a Lax equation and give  some examples.  In later sections we give nonlocal reductions of these equations.

We consider an algebra $\cal A$,  given by
\begin{equation}
{\cal A}=\left\{ \alpha= \sum\limits_{-\infty}^\infty \alpha_n(x,t)p^n \,: \alpha_n(x,t)\in Y\, \right\},
\end{equation}
where $Y$ is either $ C^\infty (S^1)$-the space of periodic functions or $S({\mathbb R})$-the spaces of smooth asymptotically  decreasing functions on $\mathbb R$.
A Poisson bracket on $\cal A$, is given by
\begin{equation}\label{bracket}
  \{f,g\}_k=p^k\left( \frac{\partial f}{\partial p}\frac{\partial g}{\partial x}-\frac{\partial f}{\partial x}\frac{\partial g}{\partial p}\right)\qquad f,g\in {\cal A},
\end{equation}
where $k\in\mathbb Z$.

On the algebra $\cal A$ we can define a trace functional
\begin{equation}
\mbox{tr}_k \, \alpha=\int_{I} \mbox{Res}_k\, \alpha\, dx  \qquad \alpha\in {\cal A},
\end{equation}
where $\mbox{Res}_k\, \alpha=\alpha_{k-1}(x,t)$. The set $I$ is either $S^1$ for the space of periodic functions or $\mathbb R$ for the space of asymptotically decaying functions.
Using the trace functional we define a non-degenerate, ad-invariant, symmetric pairing
\begin{equation}
 (\alpha,\beta)=\ \mbox{tr}_k\, \alpha\cdot \beta, \qquad \alpha,\beta\in{\cal A}.
\end{equation}
The algebra $\cal A$ admits a decomposition into sub-algebras, closed with respect to the Poisson bracket,
\begin{equation}
  {\cal A}={\cal A}_{\ge -k+1} \oplus {\cal A}_{<-k+1},
\end{equation}
where
\begin{equation}
{\cal A}_{\ge -k+1}=\left\{ \alpha_{\ge -k+1}=\sum\limits_{-k+1}^\infty \alpha_n(x,t)p^n : \alpha_n(x,t)\in Y \right\}
\end{equation}
and
\begin{equation}
{\cal A}_{ < -k+1}=\left\{\alpha_{< -k+1}= \sum\limits_{-\infty}^{-k} \alpha_n(x,t)p^n : \alpha_n(x,t)\in Y \right\}.
\end{equation}
So, we can define $r$-matrix mapping on $\cal A$ (see \cite{Bl} and references therein)
$$
R_k=\frac{1}{2}(\Pi_{\ge -k+1}-\Pi_{< -k+1})
$$
where  $\Pi_{\ge -k+1}$ and $\Pi_{< -k+1}$ are projections on the sub-algebras ${\cal A}_{\ge -k+1}$ and ${\cal A}_{ < -k+1}$ respectively.

Now  for a Lax operator
\begin{equation}
L= p^{N-1} + \sum_{i=-1}^{N-2} p^i S_i(x,t), \qquad N\in \mathbb N,
\end{equation}
we can consider a hierarchy of the  Lax equations (see \cite{GurZhel})
\begin{equation}\label{Laxeqn}
\frac{\partial L}{\partial t_n}=\left\{ R_k( L^{\frac{m}{N-1}+n}), L \right\}_k=\left\{  L^{\frac{m}{N-1}+n} _{\ge -k+1},L \right\}_k, \qquad n=1,2,\cdots,
\end{equation}
where $m=0,1,2,\cdots, (N-2)$ is fixed. This is  an integrable  hierarchy that is the Lax equations  admit multi-Hamiltonian representation (see \cite{Li}). In particular the  conserved quantities for the hierarchy are  given by
\begin{equation}\label{con_quant}
 {\cal Q}_n=\ \mbox{tr}_k L^{\frac{m}{N-1}+n}, \qquad n=1,2,\cdots \, .
\end{equation}

We note that to consider nonlocal reductions it is more convenient to write the equation \eqref{Laxeqn}  in terms of  zeros of the Lax operator.
That is we set
\begin{equation} \label{Laxfnc}
L= \frac{1}{p}\prod_{j=1}^{N}(p-u_j).
\end{equation}

Let us give some examples of the Lax equations and their conserved quantities.
First we consider examples of Lax equations corresponding to the Lax operator
\begin{equation}\label{Laxfnc_2}
\displaystyle{L=\frac{1}{p}(p-u)(p-v)}.
\end{equation}
\vspace{0.2cm}
\begin{ex}
We take the algebra $\cal A$ with $k=0$ and consider the Lax equation \eqref{Laxeqn} with a Lax operator  \eqref{Laxfnc_2}.
For  $n=2$,  we obtain the shallow water waves system
\begin{equation}\label{sww}
\begin{array}{l}
\frac{1}{2} u_{t_2}=(u+v)u_x+uv_x,\\\\
\frac{1}{2} v_{t_2}=(u+v)v_x+vu_x.
\end{array}
\end{equation}
Taking $n=3,4, \cdots$,  we can have generalized symmetries of the  shallow water waves system. For instance, the first two symmetries are
\begin{equation}\label{sww_3}
\begin{array}{l}
\frac{1}{3} u_{t_3}=(u^2 + 4uv + v^2)u_x + (2u^2 + 2uv)v_x,\\\\
 \frac{1}{3} v_{t_3}=(2v^2 + 2uv)u_x+(v^2 + 4uv + u^2)v_x,
\end{array}
\end{equation}
and
\begin{equation}\label{sww_4}
\begin{array}{l}
 \frac{1}{4}u_{t_4}= (u^3 + 9u^2v + 9uv^2 + v^3)u_x + (3u^3+9u^2v+3uv^2)v_x, \\\\
 \frac{1}{4}v_{t_4}= (3v^3+9v^2u+3vu^2)u_x + (v^3 + 9v^2u + 9vvu^2 + u^3)v_x.
\end{array}
\end{equation}
The above systems admit infinitely many conserved quantities. The first three of the conserved quantities are
\begin{eqnarray}
 &&{\cal Q}_2=\int_{I} ( u^2 v+ uv^2) \, dx,\\
&&{\cal Q}_3=\int_{I}  (u^3v + 3u^2v^2+uv^3) \, dx,\\
 &&{\cal Q}_4=\int_{I} (u^4 v+6u^3v^2 +6u^2v^3 + 6uv^4) \, dx.
\end{eqnarray}
\end{ex}

\vspace{0.2cm}
\begin{ex}
We take the algebra $\cal A$ with $k=1$ and consider the Lax equation \eqref{Laxeqn} with a Lax operator  \eqref{Laxfnc_2}.
For $n=1$  we obtain the Toda system
\begin{equation}\label{toda}
\begin{array}{l}
  u_{t_1}=uv_x,\\
  v_{t_1}=vu_x.
\end{array}
\end{equation}
Taking $n=2,3,\cdots$, we can have generalized symmetries of the  Toda system. The Toda system admits infinitely many conserved quantities. The first three of the conserved quantities are
\begin{eqnarray}
 &&{\cal Q}_1=\int_{I}  (u+v) \, dx,\\
&&{\cal Q}_2=\int_{I} (u^2 + 4uv+v^2) \, dx,\\
 &&{\cal Q}_3=\int_{I} (u^3+9u^2v+9uv^2 +v^3) \, dx.
\end{eqnarray}
\end{ex}

\noindent Let us give examples of the Lax equation with more general Lax operator.\\

\begin{ex}
We take the algebra $\cal A$ with $k=0$ and consider the Lax equation \eqref{Laxeqn} with a Lax operator
\begin{equation}\label{Laxfnc_3}
L=\frac{1}{p}(p-u)(p-v)(p-w),
\end{equation}
and $m=1$.
For $n=1$  we obtain the following system
\begin{equation}\label{systemN=3}
\begin{array}{l}
\displaystyle \frac{8}{3} u_{t_1}=(-u^2 + v^2+ w^2 + 4uv + + 4uw + 6vw)u_x +(2u^2 + 2uv + 6uw)v_x \\
\hspace {2cm} +(2u^2 + 6uv + 2uw)w_x ,\\
\displaystyle \frac{8}{3} v_{t_1}=(2v^2 + 2uv + 6vw)u_x +(-v^2 + u^2+ w^2 + 4uv + + 4vw + 6vw)v_x \\
\hspace {2cm}+(2v^2 + 6uv + 2vw)w_x ,\\
\displaystyle \frac{8}{3} w_{t_1}=(2w^2 + 6wv + 2uw)u_x ) +(2w^2 + 2wv + 6uw)v_x \\
\hspace {2cm}+(-w^2 + v^2+ u^2 + 4wv +  4uw + 6uv)w_x.
\end{array}
\end{equation}
Taking $n=2,3,\cdots$, we can have generalized symmetries of the  above system. The above system admits infinitely many conserved quantities.
The first two of the conserved quantities are
\begin{eqnarray}
 &&{\cal Q}_1=\int_{I}  ( u^2+ v^2+ w^2 - 2uv  - 2uw - 2vw ) \, dx,\\
&&{\cal Q}_2=\int_{I} (u^4+ v^4+ w^4 - 4(u^3v - u^3w - uv^3 - v^3 w - u w^3 - vw^3) + 6 (u^2 v^2+ u^2 w^2 + 6v^2w^2)\nonumber \\
 &&+ 36( u^2vw + uv^2w + uvw^2))  \, dx,
\end{eqnarray}
\end{ex}
\vspace{0.2cm}
\begin{ex}
For simplicity, we take the algebra $\cal A$ with $k=0$ and consider the Lax equation \eqref{Laxeqn} with a Lax operator
\begin{equation}
L=\frac{1}{p}(p-u)(p-u_1)(p-v)(p-v_1)
\end{equation}
and $m=0$. For $n=1$  we obtain the following system
\begin{equation}\label{systemN=4}
\begin{array}{l}
 u_{t_1}=(u_1v + u_1v_1 + v v_1)u_x +(uv +uv_1)u_{1x} +(uu_1 + uv_1)v_{x} +(uu_1 +uv)v_{1x},\\
 u_{1t_1}=(uv  + uv_1 + vv_1)u_{1x} + (u_1v +u_1v_1)u_{x} + (uu_1 +u_1v_1)v_{x} + (u_1u +u_1v)v_{1x},\\
 v_{t_1}=(uu_1+uv_1 + u_1v_1)v_{x}+(u_1v +vv_1)u_{x}+(uv +vv_1)u_{1x} +(uv +u_1v)v_{1x},\\
 v_{1t_1}=(uu_1  + uv + u_1v)v_{1x} + (u_1v_1 +vv_1)u_{x} +(uv_1 +vv_1)u_{1x} +(uv_1 +u_1v_1)v_{x}.\\
\end{array}
\end{equation}
Taking $n=2,3,\cdots$, we can have generalized symmetries of the  above system. The above system admits infinitely many conserved quantities.
The first two of the conserved quantities are
\begin{eqnarray}
&&{\cal Q}_1=\int_{I}  uu_1vv_1\, dx,\\
 &&{\cal Q}_2=\int_{I}  (u^2u_1^2v^2v_1 + u^2u_1^2vv_1^2 + u^2u_1v^2v_1^2+ uu_1^2v^2v_1^2)\, dx.
\end{eqnarray}
\end{ex}

\section{Nonlocal reductions for two component Lax operator}

To study  nonlocal reductions of the Lax equation \eqref{Laxeqn}, we start with the  case of two component systems.  That is we study the hierarchy
\begin{equation} \label{Laxeqn_N=2}
\frac{\partial L}{\partial t_n}=\left\{  L^{n} _{\ge -k+1};L \right\}_k,  \qquad n=1,2,\cdots  .
\end{equation}
with the Lax operator \eqref{Laxfnc_2}.

\subsection{Nonlocal reductions for real valued dependent variables}

In this section we assume that our dependent variables are real valued and consider  reduction of the form
\begin{equation}\label{reduction_2}
v(x,t)=\rho u(\varepsilon_1x,\varepsilon_2t)=\rho u^\varepsilon.
\end{equation}
The reduced equation is the Lax equation \eqref{Laxeqn_N=2} with the  Lax operator
\begin{equation}\label{Laxfnc_2_new}
 L=\frac{1}{p}(p-u)(p-\rho u^\varepsilon).
\end{equation}
To find admissible reductions we introduce the following change of variables
\begin{equation}
\tilde x= \varepsilon_1x,\quad \tilde t_n= \varepsilon_2t_n, \quad \tilde p= \rho p.
\end{equation}
In new  variables the Lax operator becomes
\begin{equation} \label{Laxfnc_new_2}
 L=\frac{\rho}{\tilde p}(\rho \tilde p-u^\varepsilon)(\rho \tilde p-\rho u)=\rho\tilde L,
\end{equation}
where $\displaystyle {\tilde L=\frac{1}{\tilde p}(\tilde p- u)(\tilde p - \rho u^\varepsilon)}$.
The Lax equation \eqref{Laxeqn_N=2} takes the form
\begin{equation}
\varepsilon_2 \frac{\partial}{\partial \tilde t_n}(\rho \tilde L)=
(\rho \tilde p)^k\left( \rho \frac{\partial}{\partial \tilde p}(\rho^n \tilde L_{\ge -k+1}^n)\varepsilon_1 \frac{\partial}{\partial \tilde x}(\rho \tilde L)  -
 \varepsilon_1 \frac{\partial}{\partial \tilde x}(\rho^n \tilde L_{\ge -k+1}^n)\rho \frac{\partial}{\partial \tilde p}(\rho \tilde L)  \right)
\end{equation}
or
\begin{equation}\label{Lax_eqn_red}
\rho^{k+n+1}\varepsilon_2\varepsilon_1 \frac{\partial}{\partial \tilde t_n}( \tilde L)=
 \tilde p^k\left( \frac{\partial}{\partial \tilde p}\tilde L_{\ge -k+1}^n\frac{\partial}{\partial \tilde x} \tilde L  -
 \frac{\partial}{\partial \tilde x} \tilde L_{\ge -k+1}^n \frac{\partial}{\partial \tilde p} \tilde L  \right).
\end{equation}
It is easy to see that if $\rho^{k+n+1}\varepsilon_2\varepsilon_1=1$ then the hierarchy \eqref{Laxeqn_N=2} with Lax operator \eqref{Laxfnc_2_new} is invariant under the change of variables and we have a consistent nonlocal reduction.

Since $\rho^2=1$, we have one condition $\rho^{k}\varepsilon_2\varepsilon_1=1$ for all odd $n$ and another  condition $\rho^{k+1}\varepsilon_2\varepsilon_1=1$ for all even $n$. Thus, to consider  nonlocal reduction of the whole hierarchy \eqref{Laxeqn_N=2} we have to split it into two hierarchies
\begin{equation} \label{Laxeqn_even}
\frac{\partial L}{\partial t_{2n}}=\left\{  L^{2n} _{\ge -k+1};L \right\}_k, \qquad n=1,2,\cdots
\end{equation}
and
\begin{equation} \label{Laxeqn_odd}
\frac{\partial L}{\partial t_{2n-1}}=\left\{  L^{2n-1} _{\ge -k+1};L \right\}_k, \qquad n=1,2,\cdots \, .
\end{equation}

Taking $\rho,\, \varepsilon_1$, and $\varepsilon_2$ satisfying
\begin{equation}\label{even_cond_2}
 \rho^{k+1}\varepsilon_2\varepsilon_1=1
 \end{equation}
 we obtain nonlocal reductions for the hierarchy \eqref{Laxeqn_even}. Note that these equations admit the following conserved quantities
 \begin{equation}\label{con_quant_even}
 {\cal Q}_{2n}=\ \mbox{tr}_k L^{2n}, \qquad n=1,2,\cdots\, ,
\end{equation}
 where $L$ is given by  \eqref{Laxfnc_2_new}.

Taking $\rho,\, \varepsilon_1$, and $\varepsilon_2$ satisfying
\begin{equation}\label{odd_cond_2}
\rho^{k}\varepsilon_2\varepsilon_1=1
\end{equation}
 we obtain nonlocal reductions for the hierarchy \eqref{Laxeqn_odd}. Note that these equations admit the following conserved quantities
\begin{equation}\label{con_quant_odd}
 {\cal Q}_{2n-1}=\ \mbox{tr}_k L^{2n-1}, \qquad n=1,2,\cdots\, ,
\end{equation}
 where $L$ is given by  \eqref{Laxfnc_2_new}.

\noindent Let us give some examples. As first example we consider the shallow water waves system.\\

\begin{ex}
To have nonlocal reductions of the shallow water waves system \eqref{sww}, we need to work with the hierarchy~\eqref{Laxeqn_even}.
Taking $\rho,\, \varepsilon_1$, and $\varepsilon_2$ satisfying  \eqref{even_cond_2} we obtain possible reductions.

\noindent {\bf a.} S-symmetric reduction, $\rho=-1, \varepsilon_1=-1,\varepsilon_2=1$, is
\begin{equation}
\frac{1}{2} u_{t_2}(x,t)=(u(x,t)-u(-x,t))u_x(x,t)-u(x,t)u_x(-x,t).
\end{equation}
It has infinitely many symmetries,  with the first symmetry, the reduction of \eqref{sww_4}, being
\begin{eqnarray}
&&\frac{1}{4} u_{t_4}(x,t)=(u^3(x,t) - 9u^2(x,t)u(-x,t) + 9u(x,t)u^2(-x,t) - u^3(-x,t))u_x((x,t)  \nonumber\\
&&-(3u^3(x,t)-9u^2(x,t)u(-x,t)+3u(x,t)u^2(-x,t))u_x(-x,t).
\end{eqnarray}
The above equations have infinitely many conserved quantities. The first two conserved quantities are
\begin{eqnarray}
 &&{\cal Q}_2=\int_{I}  (-u^2(x,t) u(-x,t)+ u(x,t)u^2(-x,t)) \, dx,\\
 &&{\cal Q}_4=\int_{I} (-u^4(x,t) u(-x,t)+6u^3(x,t)u^2(-x,t) -6u^2(x,t)u^3(-x,t)\nonumber\\
 &&+ u(x,t)u^4(-x,t))\, dx.
\end{eqnarray}

\noindent {\bf b.} T-symmetric reduction, $\rho=-1, \varepsilon_1=1,\varepsilon_2=-1$, is
\begin{equation}
\frac{1}{2} u_{t_2}(x,t)=(u(x,t)-u(x,-t))u_x(x,t)-u(x,t)u_x(x,-t).
\end{equation}
It has infinitely many symmetries,  with the first symmetry, the reduction of \eqref{sww_4}, being
\begin{eqnarray}
&&\frac{1}{4} u_{t_4}(x,t)=(u^3(x,t) - 9u^2(x,t)u(x,-t) + 9u(x,t)u^2(x,-t) - u^3(x,-t))u_x((x,t)  \nonumber\\
&&-(3u^3(x,t)-9u^2(x,t)u(x,-t)+3u(x,t)u^2(x,-t))u_x(x,-t).
\end{eqnarray}
The above equations have infinitely many conserved quantities. The first two conserved quantities are
\begin{eqnarray}
 &&{\cal Q}_2=\int_{I}  (-u^2(x,t) u(x,-t)+ u(x,t)u^2(x,-t)) \, dx,\\
 &&{\cal Q}_4=\int_{I} (-u^4(x,t) u(x,-t)+6u^3(x,t)u(x,-t)^2 -6u^2(x,t)u(x,-t)^3 \nonumber\\&&+ 6u(x,t)u^4(x,-t)) \, dx.
\end{eqnarray}

\noindent {\bf c.} ST-symmetric reduction, $\rho=1, \varepsilon_1=-1,\varepsilon_2=-1$, is
\begin{equation}
\frac{1}{2} u_{t_2}(x,t)=(u(x,t)+u(-x,-t))u_x(x,t)+u(x,t)u_x(-x,-t).
\end{equation}
It has infinitely many symmetries with the first symmetry, the reduction of \eqref{sww_4}, being
\begin{eqnarray}
&&\frac{1}{4} u_{t_4}(x,t)=(u^3(x,t) + 9u^2(x,t)u(-x,-t) + 9u(x,t)u^2(-x,-t) + u^3(-x,-t))u_x((x,t)  \nonumber \\
&&+(3u^3(x,t)+9u^2(x,t)u(-x,-t)+3u(x,t)u^2(-x,-t))u_x(-x,-t).
\end{eqnarray}
The above equations have infinitely many conserved quantities. The first two conserved quantities are
\begin{eqnarray}
&&{\cal Q}_2=\int_{I}  (u^2(x,t) u(-x,-t) + u^2(x,t)u(-x,-t)) \, dx,\\
&&{\cal Q}_4=\int_{I} (u^4(x,t) u(-x,-t)+6u^3(x,t)u(-x,-t)^2 +6u^2(x,t)u^3(-x,-t) \nonumber\\&&+ u(x,t)u^4(-x,-t)) \, dx.
\end{eqnarray}
\end{ex}

\noindent We can also  consider the reductions for the odd symmetries of the shallow water waves equations.\\

\begin{ex}
To have nonlocal reductions of the  system \eqref{sww_3}, we need to work with the hierarchy \eqref{Laxeqn_odd}.
Taking $\rho,\, \varepsilon_1$, and $\varepsilon_2,$ satisfying  \eqref{odd_cond_2} we obtain one possible  reduction.

\noindent ST-symmetric reduction, $\rho=1, \varepsilon_1=-1,\varepsilon_2=-1$, is
\begin{eqnarray}
&&\frac{1}{3} u_{t_3}(x,t)=(u(x,t)^2 + 4u(x,t)u(-x,-t) + u^2(-x,-t))u_x(x,t) \nonumber \\
&&+(2u^2(x,t) + 2u(x,t)u(-x,-t))u_x(-x,-t).
\end{eqnarray}

The above system has infinitely many symmetries and conserved quantities. The first two conserved quantities are
\begin{eqnarray}
&&{\cal Q}_3=\int_{I}  (u^3(x,t)u(-x,-t) + 3 u^2(x,t)u^2(-x,-t)+u(x,t)u(-x,-t)^3) \, dx, \\
&&{\cal Q}_5=\int_{I}  (u^5(x,t)u(-x,-t) + 10 u^4(x,t) u^2(-x,-t)+20 u^3(x,t) u^3(-x,-t) \nonumber\\
&& + 10 u^2(x,t) u^4(-x,-t)+ u(x,t)u^5(-x,-t)) \, dx.
\end{eqnarray}
\end{ex}

\noindent Let us also consider the reductions of the Toda system.\\

\begin{ex}
To have nonlocal reductions of the Toda system \eqref{toda}, we need to work with the hierarchy \eqref{Laxeqn_odd}.
Taking $\rho,\, \varepsilon_1,$ and $\varepsilon_2$ satisfying  \eqref{odd_cond_2} we can obtain the following reductions: S-symmetric reduction, $\rho=-1, \varepsilon_1=-1,\varepsilon_2=1$; T-symmetric reduction,
 $\rho=-1, \varepsilon_1=1,\varepsilon_2=-1$;  ST-symmetric reduction, $\rho=1, \varepsilon_1=-1,\varepsilon_2=-1$.
For instance, the ST-symmetric reduction of the Toda system is
 \begin{equation}
 u_{t_1}(x,t)=u(x,t)u_x(-x,-t). \\
\end{equation}
It has infinitely many  symmetries,  with the first symmetry
\begin{eqnarray}
&&\frac{1}{3} u_{t_3}(x,t)= 2u(x,t) \bigg(u(x,t)u(-x,-t) + u^2(-x,-t)\bigg)u_x(x,t)\nonumber  \\
  &&+ u(x,t)\bigg(u^2(x,t) + 4u(x,t)u(-x,-t) + u^2(x,t)\bigg)u_x(-x,-t),
\end{eqnarray}
which is the  ST-symmetric  reduction of the second equation in the hierarchy \eqref{Laxeqn_N=2}.
It also  has infinitely many conserved quantities. The first two conserved quantities are
\begin{eqnarray}
&&{\cal Q}_1=\int_{I}  (u(x,t)+ u(-x,-t))  \, dx,\\
&&{\cal Q}_3=\int_{I} (u^3(x,t) +9 u^2(x,t) u(-x,-t)
  + 9 u(x,t) u^2(-x,-t) - u^3(-x,-t)) \, dx .
\end{eqnarray}
\end{ex}
\vspace{0.2cm}
\begin{rem}
We considered the reductions for the whole hierarchy \eqref{Laxeqn_N=2}. We can also consider nonlocal reductions just for  one equation from the hierarchy. For instance, taking $k=0$ and  $n=2$ in  \eqref{Laxeqn_N=2} we
get the following equations:
\begin{align}
& \frac{1}{2} u_{t_2}=(u+v)u_x+uv_x,\\
& \frac{1}{2} v_{t_2}=(u+v)v_x+vu_x.
\end{align}
Let us consider nonlocal reductions of these equations only.
Assume that $v(x,t)=\rho u(\varepsilon_1x,\varepsilon_2t)=\rho u^{\varepsilon}$, where $\varepsilon_1^2=\varepsilon_2^2=1$, and $\rho$ is a real constant, then the above system takes form
\begin{align}
& \frac{1}{2\rho} u_t=(u^{\varepsilon}+\frac{1}{\rho}u )u_x+ uu_x^{\varepsilon},\\
& \frac{1}{2}\varepsilon_1\varepsilon_2 u_t=(u^{\varepsilon}+\rho u)u_x+uu_x^{\varepsilon}.
\end{align}
For consistency, we must have $\rho=\varepsilon_1\varepsilon_2$ which is the same condition as equality \eqref{even_cond_2}. Therefore working  with the whole hierarchy we obtain all possible cases.
\end{rem}

\subsection{Nonlocal reductions for complex  valued dependent variables}

In this section we assume that all dependent variables  are complex valued functions. To study reductions of such equations it is convenient to introduce a constant in the Lax equation  \eqref{Laxeqn_N=2} by setting new time $t'=at$, $a\in {\mathbb C}$. So we consider a hierarchy
\begin{equation} \label{Laxeqn_N=2_com}
a\frac{\partial L}{\partial t_n}=\left\{  L^{n} _{\ge -k+1};L \right\}_k,  \qquad n=1,2,\cdots .
\end{equation}
Note that after the change of the time  variable we shall use again $t$ to denote new time.

The reduction is of the form
\begin{equation}
v(x,t)=\rho \bar u(\varepsilon_1x,\varepsilon_2t)=\rho \bar{u}^{\varepsilon}.
\end{equation}
To find admissible reductions we use the following change of variables
\begin{equation}
\tilde x= \varepsilon_1x,\quad \tilde t_n= \varepsilon_2t_n, \quad \tilde p= \rho p.
\end{equation}
In new  variables the Lax operator becomes
\begin{equation}
 L=\frac{\rho}{\tilde p}(\rho \tilde p- u^\varepsilon)(\rho \tilde p-\rho \bar u)=\rho \bar L',
\end{equation}
where $\displaystyle {L' =\frac{1}{\tilde p}(\tilde p- u)(\tilde p - \rho \bar u^\varepsilon)}$ and
the Lax equation \eqref{Laxeqn_N=2} takes the form
\begin{equation}
a\varepsilon_2 \frac{\partial}{\partial \tilde t_n}(\rho \bar L')=
(\rho \tilde p)^k\left( \rho \frac{\partial}{\partial \tilde p}(\rho^n  (\bar L')_{\ge -k+1}^n)\varepsilon_1 \frac{\partial}{\partial \tilde x}(\rho \bar L')  -
 \varepsilon_1 \frac{\partial}{\partial \tilde x}(\rho^n  (\bar L')_{\ge -k+1}^n)\rho \frac{\partial}{\partial \tilde p}(\rho \bar L')  \right).
\end{equation}
Taking complex conjugates of both sides in the above equality we can get
\begin{equation}
\overline{a}\rho^{k+n+1}\varepsilon_2\varepsilon_1 \frac{\partial}{\partial \tilde t_n}(  L')=
 \tilde p^k\left( \frac{\partial}{\partial \tilde p} (L')_{\ge -k+1}^n\frac{\partial}{\partial \tilde x}  L'  -
 \frac{\partial}{\partial \tilde x}  (L')_{\ge -k+1}^n \frac{\partial}{\partial \tilde p} L'  \right).
\end{equation}
If $\overline{a}\rho^{k+n+1}\varepsilon_2\varepsilon_1=a$ then the hierarchy \eqref{Laxeqn_N=2_com} is invariant under the change of variables and we have a consistent nonlocal reduction.

Since $\rho^2=1$ we have one condition $\overline{a}\rho^{k}\varepsilon_2\varepsilon_1=a$ for all odd $n$ and another  condition $\overline{a}\rho^{k+1}\varepsilon_2\varepsilon_1=a$ for all even $n$. Thus, to consider  local reduction of the hierarchy \eqref{Laxeqn_N=2_com} we have to split it into two hierarchies \eqref{Laxeqn_even} and \eqref{Laxeqn_odd} as in the real case.

\noindent Let us give  one example of the reduction for complex valued dependent variables.\\

\begin{ex}
We take $k=0$ and consider the hierarchy \eqref{Laxeqn_N=2_com} with $n$ even.
The first  system corresponding to $n=2$ is
\begin{equation}
\begin{array}{l}
a u_{t_2}=2(u+v)u_x+2uv_x,\\
a v_{t_2}=2(u+v)v_x+2vu_x.
\end{array}
\end{equation}
Taking $\rho,\, \varepsilon_1,$ and $\varepsilon_2$ satisfying  $\overline{a}\rho\varepsilon_2\varepsilon_1=a$   we  obtain possible reductions.\\

\noindent
{\bf a.} S-symmetric reduction, $\varepsilon_1=-1, \, \varepsilon_2=1$, is
\begin{equation}
\frac{a}{2} u_t(x,t)=(u(x,t)+\rho \bar{u}(-x,t))u_x(x,t)+\rho u(x,t)\bar{u}_x(-x,t),
\end{equation}
where we can take $\rho=1$ and the constant $a$ equal to a pure imaginary number or $\rho=-1$ and the constant $a$ equal to a real number. The above equation has infinitely many symmetries and conserved quantities.
The first two conserved quantities are
\begin{eqnarray}
 &&{\cal Q}_2=\int_{I}  (-u^2(x,t) \rho\bar u(-x,t)+ u(x,t)\rho\bar u^2(-x,t)) \, dx,\\
&&{\cal Q}_4=\int_{I} (-u^4(x,t) \rho\bar u(-x,t)+6u^3(x,t) \rho\bar u^2(-x,t) -6u^2(x,t)\rho\bar u^3(-x,t)\nonumber
\\&&+ u(x,t)\rho\bar u^4(-x,t))\, dx.
\end{eqnarray}

\noindent
{\bf b.} T-symmetric  reduction, $\varepsilon_1=1, \, \varepsilon_2=-1$, is
\begin{equation}
\frac{a}{2} u_t(x,t)=(u(x,t)+\rho \bar{u}(x,-t))u_x(x,t)+ \rho u(x,t)\bar{u}_x(x,-t),
\end{equation}
where we can take  $\rho=1$ and the constant $a$ equal to a pure imaginary number  or $\rho=-1$  and the constant $a$ equal to a real number.
The above equation has infinitely many symmetries and conserved quantities. The first two conserved quantities are
\begin{eqnarray}
&& {\cal Q}_2=\int_{I}  (-u^2(x,t) \rho\bar u(x,-t)+ u(x,t) \rho\bar u^2(x,-t)) \, dx,\\
&&{\cal Q}_4=\int_{I} (-u^4(x,t) \rho\bar u(x,-t)+6u^3(x,t) \rho\bar u^2(x,-t) -6u^2(x,t)\rho \bar u^3(x,-t)\nonumber\\
 &&+ u(x,t)\rho\bar u^4(x,-t))\, dx.
\end{eqnarray}

\noindent
{\bf c.} ST-symmetric reduction, $\varepsilon_1=-1, \, \varepsilon_2=-1$, is
\begin{equation}
\frac{a}{2} u_t(x,t)=(u(x,t)+ \rho\bar{u}(-x,-t))u_x(x,t)+\rho u(x,t)\bar{u}_x(-x,-t),
\end{equation}
where we can take $\rho=1$ and the constant $a$ equal to a real number or $\rho=-1$ and the constant $a$ equal to a pure imaginary number.
The above equation has infinitely many symmetries and conserved quantities. The first two conserved quantities are
\begin{eqnarray}
 &&{\cal Q}_2=\int_{I}  (-u^2(x,t) \rho\bar u(-x,-t)+ u(x,t) \rho\bar u^2(-x,-t)) \, dx,\\
&&{\cal Q}_4=\int_{I} (-u^4(x,t) \rho\bar u(-x,-t)+6u^3(x,t) \rho\bar u^2(-x,-t) -6u^2(x,t) \rho\bar u^3(-x,-t)\nonumber\\
 &&+ u(x,t)\rho\bar u^4(-x,-t))\, dx.
\end{eqnarray}
\end{ex}

\section{Reductions with a general Lax operator}

To study  nonlocal reductions of the Lax equation \eqref{Laxeqn} with the general Lax operator \eqref{Laxfnc}, we can divide the variables $u_i$ into pairs and set variables that do not have a pair to zero. For each pair of variables we consider a relation similar to \eqref{reduction_2} and proceed as before. As an example let us consider two special cases.

\subsection{Three component Lax operator}
In this section we take the Lax operator
\begin{equation}
\displaystyle{L=\frac{1}{p}(p-u)(p-v)(p-w)},
\end{equation}
and  consider the hierarchy \eqref{Laxeqn} with $m=1$. Thus we have
\begin{equation} \label{Laxeqn_N=3}
\frac{\partial L}{\partial t_n}=\left\{  L^{\frac{1}{2}+n}_{\ge -k+1};L \right\}_k,  \qquad n=1,2,\cdots .
\end{equation}

\subsubsection {Real valued dependent variables}
First we study reductions for real valued dependent variables.
A nonlocal reduction has a form
\begin{equation}\label{reduction_3}
v(x,t)=\rho u(\varepsilon_1x,\varepsilon_2t)=\rho u^\varepsilon.
\end{equation}
 To find admissible reductions we use the following change of variables
\begin{equation}
\tilde x= \varepsilon_1x,\quad \tilde t_n= \varepsilon_2t_n, \quad \tilde p= \rho p.
\end{equation}
In new  variables the Lax operator becomes
\begin{equation}
 L=\frac{1}{\rho \tilde p}(\rho \tilde p-u^\varepsilon)(\rho \tilde p-\rho u)(\rho \tilde p-w^\varepsilon)=\tilde L,
\end{equation}
where $\tilde L=\displaystyle {\frac{1}{ \tilde p}(\tilde p - u)(\tilde p-\rho u^\varepsilon)(\tilde p-\rho w^\varepsilon)}$. Also in new variables we have $L^{\frac{1}{2}}=\rho \tilde L^{\frac{1}{2}}$.

The Lax equation \eqref{Laxeqn_N=3} takes the form
\begin{equation}
\varepsilon_2 \frac{\partial}{\partial \tilde t_n}\tilde L=
(\rho \tilde p)^k\left( \rho\frac{\partial}{\partial \tilde p}\rho \tilde L^{\frac{1}{2}+n}_{\ge -k+1}\varepsilon_1 \frac{\partial}{\partial \tilde x} \tilde L  -
 \varepsilon_1 \frac{\partial}{\partial \tilde x} \rho \tilde L^{\frac{1}{2}+n}_{\ge -k+1}\rho \frac{\partial}{\partial \tilde p} \tilde L  \right),
\end{equation}
or
\begin{equation}\label{Laxeqn_N=3_tilde_L}
\rho^{k}\varepsilon_2\varepsilon_1 \frac{\partial}{\partial \tilde t_n}\tilde L=
 \tilde p^k\left( \frac{\partial}{\partial \tilde p}\tilde L_{\ge -k+1}^{\frac{1}{2}+n}\frac{\partial}{\partial \tilde x} \tilde L  -
 \frac{\partial}{\partial \tilde x} \tilde L_{\ge -k+1}^{\frac{1}{2}+n} \frac{\partial}{\partial \tilde p} \tilde L  \right).
\end{equation}
Now we set $w=0$ in the equation \eqref{Laxeqn_N=3} and $w^\varepsilon=0$ in the equation \eqref{Laxeqn_N=3_tilde_L}. It is easy to see that under these conditions the hierarchy \eqref{Laxeqn_N=3} is invariant under the change of variables if
\begin{equation}\label{cond_3}
\rho^{k}\varepsilon_2\varepsilon_1=1.
\end{equation}
So, we have a consistent nonlocal reduction if the above condition is satisfied.\\

\begin{ex}
Let us consider reductions for the system \eqref{systemN=3}.  The condition \eqref{cond_3} takes form $\varepsilon_2\varepsilon_1=1$. Thus we can have only
$ST$-symmetric reduction. Taking $\varepsilon_1=-1$ and $\varepsilon_2=-1$ we get $ST$-symmetric  nonlocal reduction.
For $\rho=1$, the reduced equation is
\begin{eqnarray}
 &&\frac{8}{3} u_{t_1}=(-u^2(x,t) + u^2(-x,-t) + 4u(x,t)u(-x,-t))u_x(x,t) \nonumber\\
&& +(2u^2(x,t)+ 2u(x,t)u(-x,-t))u_x (-x,-t).
\end{eqnarray}
The above system has infinitely many generalized symmetries and conserved quantities. The first two conserved quantities are
\begin{eqnarray}
 &&{\cal Q}_1=\int_{I}  (u^2(x,t)+ u^2(-x,-t)-2u(x,t)u(-x,-t)) \, dx,\\
&&{\cal Q}_2=\int_{I} (u^4(x,t)+ u^4(-x,-t) -4u^3(x,t)u(-x,-t) - 4u(x,t)u^3(-x,-t) \nonumber\\&&+ 6 u^2(x,t) u^2(-x,-t))  \, dx.
\end{eqnarray}
\end{ex}

\subsubsection {Complex valued dependent variables}

In this section we assume that all variables  in the Lax equation \eqref{Laxeqn_N=3} are complex valued functions. As in the case $N=2$, we transform the Lax equation to the form
\begin{equation} \label{Laxeqn_N=3_com}
a\frac{\partial L}{\partial t_n}=\left\{  L^{\frac{1}{2}+n}_{\ge -k+1};L \right\}_k,  \qquad n=1,2,\cdots .
\end{equation}

The reduction is of the form
\begin{equation}
v(x,t)=\rho \bar u(\varepsilon_1x,\varepsilon_2t)=\rho \bar{u}^{\varepsilon}.
\end{equation}
To find admissible reductions we use the following change of variables
\begin{equation}
\tilde x= \varepsilon_1x,\quad \tilde t_n= \varepsilon_2t_n, \quad \tilde p= \rho p,
\end{equation}
for the Lax equation~\eqref{Laxeqn_N=3_com}.

In new  variables the Lax operator becomes
\begin{equation}
 L=\frac{1}{\rho \tilde p}(\rho \tilde p-u^\varepsilon)(\rho \tilde p-\rho \bar u)(\rho \tilde p-w^\varepsilon)=\bar L',
\end{equation}
where $ L'=\displaystyle {\frac{1}{ \tilde p}(\tilde p - u)(\tilde p-\rho \bar u^\varepsilon)(\tilde p-\rho \bar w^\varepsilon)}$.
Also in new variables we have $ L^{\frac{1}{2}}=\rho \bar {L'}^{\frac{1}{2}}$.

The Lax equation \eqref{Laxeqn_N=3_com} takes the form
\begin{equation}
a\varepsilon_2 \frac{\partial}{\partial \tilde t_n}\bar L'=
(\rho \tilde p)^k\left( \rho \frac{\partial}{\partial \tilde p} \rho \bar {L'}_{\ge -k+1}^{\frac{1}{2}+n} \varepsilon_1 \frac{\partial}{\partial \tilde x} \bar L'  -
 \varepsilon_1 \frac{\partial}{\partial \tilde x} \rho \bar {L'}_{\ge -k+1}^{\frac{1}{2}+n} \rho \frac{\partial}{\partial \tilde p} \bar L'  \right).
\end{equation}
Taking complex conjugates of both sides in the above equality we get
\begin{equation}
\overline{a}\rho^{k+1}\varepsilon_2\varepsilon_1 \frac{\partial}{\partial \tilde t_n}  L'=
 \tilde p^k\left( \frac{\partial}{\partial \tilde p} {L'}_{\ge -k+1}^{\frac{1}{2}+n}\frac{\partial}{\partial \tilde x}  L'  -
 \frac{\partial}{\partial \tilde x}  {L'}_{\ge -k+1}^{\frac{1}{2}+n} \frac{\partial}{\partial \tilde p} L'  \right)
\end{equation}
Now we set $w=0$ in the equation \eqref{Laxeqn_N=3} and $w^\varepsilon=0$ in the equation \eqref{Laxeqn_N=3_tilde_L}. It is easy to see that under these conditions the hierarchy \eqref{Laxeqn_N=3} is invariant under the change of variables if
\begin{equation}\label{cond_3_com}
\overline{a}\rho^{k}\varepsilon_2\varepsilon_1=a.
\end{equation}
So, we have a consistent nonlocal reduction if the above condition is satisfied.\\

\begin{ex}
Let us consider reductions for the modified system \eqref{systemN=3} (the time is redefined to introduce the multiple $a$).
The condition \eqref{cond_3_com} takes the form $\bar a \varepsilon_2\varepsilon_1=a$. Thus we can have $S$-symmetric, $T$-symmetric, and $ST$-symmetric reductions.
For instance, taking $a$ purely imaginary, $\varepsilon_1=-1$, and $\varepsilon_2=1$ we get $S$-symmetric  nonlocal reduction. For $\rho=1$, the reduced equation is
\begin{eqnarray}
&&\frac{8}{3} a u_{t_1}=(-u^2(x,t) + \bar u^2(-x,t) + 4u(x,t)\bar u(-x,t))u_x(x,t)\nonumber \\
&&+(2u^2(x,t) + 2u(x,t)\bar u(-x,t))\bar u_x (-x,t).
\end{eqnarray}
The above system has infinitely many generalized symmetries and conserved quantities. The first two conserved quantities are
\begin{eqnarray}
 &&{\cal Q}_1=\int_{I}  (u^2(x,t)+ \bar u^2(-x,t)- 2u(x,t)\bar u(-x,t)) \, dx,\\
&&{\cal Q}_2=\int_{I} (u^4(x,t)+ u^4(-x,t) - 4u^3(x,t)\bar u(-x,t) - 4u(x,t)\bar u^3(-x,t)\nonumber \\&&+ 6 u^2(x,t) \bar u^2(-x,t) ) \, dx.
\end{eqnarray}
\end{ex}

\subsection{Four component Lax operator }

In this section we take the Lax operator
\begin{equation}\label{Laxfnc_4}
\displaystyle{L=\frac{1}{p}(p-u)(p-u_1)(p-v)(p-v_1)},
\end{equation}
and  consider the hierarchy \eqref{Laxeqn} with $m=0$, for simplicity. Thus we have
\begin{equation} \label{Laxeqn_N=4}
\frac{\partial L}{\partial t_n}=\left\{  L^n _{\ge -k+1};L \right\}_k,  \qquad n=1,2,\cdots .
\end{equation}

\subsubsection {Real valued dependent variables}
First we study reductions for real valued dependent variables.
A nonlocal reduction has a form
\begin{equation}\label{reduction_4}
u_1(x,t)=\rho u(\varepsilon_1x,\varepsilon_2t)=\rho u^{\varepsilon}, \quad v_1(x,t)=\rho v(\varepsilon_1x,\varepsilon_2t)=\rho v^{\varepsilon}.
\end{equation}
To find admissible reductions we use the following change of variables
\begin{equation}
\tilde x= \varepsilon_1x,\quad \tilde t_n= \varepsilon_2t_n, \quad \tilde p= \rho p.
\end{equation}
In new  variables the Lax operator becomes
\begin{equation}
 L=\frac{1}{\rho\tilde p}(\rho \tilde p-u^\varepsilon)(\rho \tilde p-\rho u)(\rho \tilde p-v^\varepsilon)(\rho \tilde p-\rho v)=\rho\tilde L,
\end{equation}
where $\displaystyle {\tilde L=\frac{1}{\tilde p}(\tilde p- u)(\tilde p - \rho u^\varepsilon)(\rho \tilde p-v^\varepsilon)(\rho \tilde p-\rho v)}$.
The Lax equation \eqref{Laxeqn_N=4} takes the form
\begin{equation}
\varepsilon_2 \frac{\partial}{\partial \tilde t_n}(\rho \tilde L)=
(\rho \tilde p)^k\left( \rho\frac{\partial}{\partial \tilde p}(\rho^n\tilde L^n)_{\ge -k+1}\varepsilon_1 \frac{\partial}{\partial \tilde x}( \tilde L)  -
 \varepsilon_1 \frac{\partial}{\partial \tilde x} (\rho^n\tilde L^n)_{\ge -k+1}\rho \frac{\partial}{\partial \tilde p}( \tilde L)  \right)
\end{equation}
or
\begin{equation}
\rho^{n+k+1}\varepsilon_2\varepsilon_1 \frac{\partial}{\partial \tilde t_n}( \tilde L)=
 \tilde p^k\left( \frac{\partial}{\partial \tilde p}\tilde L_{\ge -k+1}^n\frac{\partial}{\partial \tilde x} \tilde L  -
 \frac{\partial}{\partial \tilde x} \tilde L_{\ge -k+1}^n \frac{\partial}{\partial \tilde p} \tilde L  \right).
\end{equation}
If
\begin{equation}\label{cond_4}
\rho^{n+k+1}\varepsilon_2\varepsilon_1=1
\end{equation}
then the hierarchy \eqref{Laxeqn_N=4} is invariant under change of variables and we have a consistent nonlocal reduction.
As in the case $N=2$, to have integrable hierarchy of nonlocal systems we have to split our hierarchy into two hierarchies:
one with even $n$ and another with odd $n$.\\

\begin{ex}
Let us consider reductions for the system \eqref{systemN=4}. It follows that for nonlocal reductions we have to consider the systems of hierarchy \eqref{Laxfnc_4},
corresponding to odd $n$. The condition \eqref{cond_4} takes the form $\varepsilon_2\varepsilon_1=1$. Thus we can have only $ST$-symmetric reduction, $\varepsilon_1=-1,\varepsilon_2=-1$.
Taking $\rho=1$ and setting $u_1(x,t)= u(-x,-t)=u^\varepsilon$ and $ v_1(x,t)=v(-x,-t)=v^\varepsilon$
we obtain ST-symmetric nonlocal reduction.
\begin{equation}
\begin{array}{l}
 u_{t_1}=(u^\varepsilon v + u^\varepsilon v^\varepsilon + v v^\varepsilon)u_x +(uv +uv^\varepsilon)u^\varepsilon_{x} +(uu^\varepsilon + uv^\varepsilon)v_{x} +(uu^\varepsilon +uv)v^\varepsilon_{x}\\
 v_{t_1}=(uu^\varepsilon+uv^\varepsilon + u^\varepsilon v^\varepsilon)v_{x}+(u^\varepsilon v +vv^\varepsilon)u_{x}+(uv +vv^\varepsilon)u^\varepsilon_{x} +(uv +u^\varepsilon v)v^\varepsilon_{x}.
\end{array}
\end{equation}
The above system has infinitely many generalized symmetries and conserved quantities. The first two conserved quantities are
\begin{eqnarray}
&&{\cal Q}_1=\int_{I}  uu^\varepsilon vv^\varepsilon \, dx,\\
&&{\cal Q}_2=\int_{I}  (u^2 (u^\varepsilon)^2v^2v^\varepsilon + u^2(u^\varepsilon)^2v(v^\varepsilon)^2+u^2u^\varepsilon v^2 (v^\varepsilon)^2+ u(u^\varepsilon)^2v^2(v^\varepsilon)^2)\, dx.
\end{eqnarray}
\end{ex}

\subsubsection {Complex valued dependent variables}

In this section we assume that all variables  in the Lax equation \eqref{Laxeqn_N=4} are complex valued functions. As in the case $N=2$, we transform the Lax equation to the form
\begin{equation} \label{Laxeqn_N=4_com}
a\frac{\partial L}{\partial t_n}=\left\{  L^{n} _{\ge -k+1};L \right\}_k,  \qquad n=1,2,\cdots .
\end{equation}

The reduction is of the form
\begin{equation}
u_1(x,t)=\rho \bar u(\varepsilon_1x,\varepsilon_2t)=\rho \bar{u}^{\varepsilon}, \qquad v_1(x,t)=\rho \bar v(\varepsilon_1x,\varepsilon_2t)=\rho \bar{v}^{\varepsilon}.
\end{equation}
To find admissible reductions we use the following change of variables
\begin{equation}
\tilde x= \varepsilon_1x,\quad \tilde t_n= \varepsilon_2t_n, \quad \tilde p= \rho p,
\end{equation}
for the Lax equation~\eqref{Laxeqn_N=4_com}.

In new  variables the Lax operator becomes
\begin{equation}
 L=\frac{\rho}{\tilde p}(\rho \tilde p- u^\varepsilon)(\rho \tilde p-\rho \bar u)(\rho \tilde p- v^\varepsilon)(\rho \tilde p-\rho \bar v)=\rho \bar L',
\end{equation}
where $\displaystyle {L' =\frac{1}{\tilde p}(\tilde p- u)(\tilde p - \rho \bar u^\varepsilon)(\tilde p- v)(\tilde p - \rho \bar v^\varepsilon)}$.
The Lax equation \eqref{Laxeqn_N=4_com} takes the form
\begin{equation}
a\varepsilon_2 \frac{\partial}{\partial \tilde t_n}(\rho \bar L')=
(\rho \tilde p)^k\left( \rho \frac{\partial}{\partial \tilde p}(\rho^n  (\bar L')^n)_{\ge -k+1}\varepsilon_1 \frac{\partial}{\partial \tilde x}(\rho \bar L')  -
 \varepsilon_1 \frac{\partial}{\partial \tilde x}(\rho^n  (\bar L')^n)_{\ge -k+1}\rho \frac{\partial}{\partial \tilde p}(\rho \bar L')  \right).
\end{equation}
Taking complex conjugates of both sides in the above equality we get
\begin{equation}
\overline{a}\rho^{k+n+1}\varepsilon_2\varepsilon_1 \frac{\partial}{\partial \tilde t_n}(  L')=
 \tilde p^k\left( \frac{\partial}{\partial \tilde p} (L')_{\ge -k+1}^n\frac{\partial}{\partial \tilde x}  L'  -
 \frac{\partial}{\partial \tilde x}  (L')_{\ge -k+1}^n \frac{\partial}{\partial \tilde p} L'  \right).
\end{equation}
If
\begin{equation}\label{cond_4_com}
\overline{a}\rho^{k+n+1}\varepsilon_2\varepsilon_1=a
\end{equation}
then the hierarchy \eqref{Laxeqn_N=4_com} is invariant under the change of variables and we have a consistent nonlocal reduction.

Since $\rho^2=1$ we have one condition $\overline{a}\rho^{k}\varepsilon_2\varepsilon_1=a$ for all odd $n$ and another  condition $\overline{a}\rho^{k+1}\varepsilon_2\varepsilon_1=a$ for all even $n$. Thus, to consider  nonlocal reduction of the hierarchy \eqref{Laxeqn_N=2_com} we have to split it into two hierarchies as in the real case.\\

\begin{ex}
Let us consider reductions for the modified system \eqref{systemN=4} (the time is redefined to introduce the multiple $a$). It follows that for nonlocal reductions we have to consider the systems of hierarchy \eqref{Laxeqn_N=4_com} corresponding to odd $n$. The condition \eqref{cond_4_com} takes form $\bar a\varepsilon_2\varepsilon_1=a$. Thus we can have $S$-symmetric, $T$-symmetric and $ST$-symmetric reduction.
For instance, taking $a$ purely imaginary and $\varepsilon_1=-1$ ,$\varepsilon_2=1$ we get $S$-symmetric  nonlocal reduction. Taking $\rho=1$ and  setting $ u_1(x,t)=\bar u(-x,t)=\bar{u}^\varepsilon$ and $v_1(x,t)=\bar v(-x,t)=\bar{v}^\varepsilon$  we  can write the reduced equation as
\begin{equation}
\begin{array}{l}
 au_{t_1}=(\bar{u}^\varepsilon v + \bar{u}^\varepsilon \bar{v}^\varepsilon + v \bar{v}^\varepsilon)u_x +(uv +u\bar{v}^\varepsilon)\bar{u}^\varepsilon_{x} +(u\bar{u}^\varepsilon + u\bar{v}^\varepsilon)v_{x} +(u\bar{u}^\varepsilon +uv)\bar{v}^\varepsilon_{x},\\
 av_{t_1}=(u\bar{u}^\varepsilon+u\bar{v}^\varepsilon + \bar{u}^\varepsilon \bar{v}^\varepsilon)v_{x}+(\bar{u}^\varepsilon v +v\bar{v}^\varepsilon)u_{x}+(uv +v\bar{v}^\varepsilon)\bar{u}^\varepsilon_{x} +(uv +\bar{u}^\varepsilon v)\bar{v}^\varepsilon_{x}.
\end{array}
\end{equation}
The above system has infinitely many generalized symmetries and conserved quantities. The first two conserved quantities are
\begin{eqnarray}
&&{\cal Q}_1=\int_{I}  u\bar{u}^\varepsilon v\bar{v}^\varepsilon \, dx,\\
&&{\cal Q}_2=\int_{I}  (u^2 (\bar{u}^\varepsilon)^2v^2\bar{v}^\varepsilon + u^2(\bar{u}^\varepsilon)^2v(\bar{v}^\varepsilon)^2+u^2\bar{u}^\varepsilon v^2 (\bar{v}^\varepsilon)^2+ u(\bar{u}^\varepsilon)^2v^2(\bar{v}^\varepsilon)^2)\, dx.
\end{eqnarray}
\end{ex}

\section{Conclusion}
We have found all possible nonlocal reductions of the hydrodynamic type of equations in $1+1$ dimensions. Such reductions are possible only for all even numbered systems. We have given examples of nonlocal reductions for $N=2$ and $N=4$ when the  dynamical variables are real and complex valued functions. For the complex valued dynamical variables we have all kinds of nonlocal reductions, namely space reflection, time reflection, and space-time reflections. The new nonlocal systems admit Lax representations and possess infinitely many conserved quantities. We gave some examples of these conserved quantities of the new nonlocal reduced systems.

\section{Acknowledgment}
  This work is partially supported by the Scientific
and Technological Research Council of Turkey (T\"{U}B\.{I}TAK).\\

\end{document}